# Electric generation from drops impacting onto charged surfaces

Hao Wu[1,2,3*], Niels Mendel[3], Dirk van den Ende[3], Guofu Zhou[1,2,4*], Frieder Mugele[3*]

[1] Guangdong Provincial Key Laboratory of Optical Information Materials and Technology & Institute of Electronic Paper Displays, South China Academy of Advanced Optoelectronics, South China Normal University, Guangzhou 510006, P. R. China

[2] National Center for International Research on Green Optoelectronics, South China Normal University, Guangzhou 510006, P. R. China

[3] Physics of Complex Fluids, Faculty of Science and Technology, MESA+ Institute for Nanotechnology, University of Twente, Enschede 7500AE, the Netherlands

[4] Shenzhen Guohua Optoelectronics Tech. Co. Ltd., Shenzhen 518110, P. R. China

[*] Emails: haowu.ut@gmail.com (H.W.); guofu.zhou@m.scnu.edu.cn (G.Z.); f.mugele@utwente.nl (F.M.)

**Abstract**

The impact of liquid drops onto solid surfaces leads to conversion of kinetic energy of directed drop motion into various forms of energy including surface energy, vibrational energy, heat, and – under suitable conditions – electrical energy. The latter has attracted substantial attention in recent years for its potential to directly convert energy from random environmental flows such as rainfall, spray, and wave motion on the sea to electrical energy. Despite the invention of numerous configurations of such energy harvesters, the underlying physical principles and optimum operation conditions have remained elusive. In this letter, we use a combination of high-speed electrical current and video imaging measurements to develop a parameter-free quantitative description of the energy harvesting process for an optimized electrode configuration. A novel electrowetting-assisted charge injection method, EWCI, enables highly stable surface charges and robust energy conversion for several months with record efficiencies exceeding 2.5% of the initial kinetic energy.

Like the generation of mechanical drop motion by electrical actuation in electrowetting[1,2], the inverse process of generating electrical signals from mechanical motion arises from variations of the capacity between one or more fixed electrodes and the electrically conductive mobile and deformable drop, which acts as the second electrode[3,4]. Relying on a facile motion of aqueous drops, both types of processes require hydrophobic surfaces. To induce an electrical current, a potential difference needs to be present between the fixed electrode(s) and the deformed drop. This can be achieved by an external power supply [3] (at the expense of some electrical losses) or – more elegantly – by an intrinsic charge transfer process between the moving drop and the surface, generally denoted as tribo-charging[5-7]. While ubiquitous, tribo-charging notoriously depends on process conditions, fluid composition and the specific solid material, which is probably related to general problem of heavily discussed spontaneous charge generation at hydrophobic-water interfaces[8-11]. All this has hampered a quantitative analysis of the energy harvesting process and thus a systematic optimization beyond the realization of the benefits of higher intrinsic charge densities[12].

In our experiments, we release millimeter-sized drops from a height $h$ of 4 to 7 cm and simultaneously monitor the drop-substrate interfacial area through the transparent substrate and the electrical current through an external load resistor $R_L$ as they fall onto micrometer thin amorphous fluoropolymer (AFP) films covering a submerged homogeneous electrode, Fig. 1a. The AFP films are pre-charged prior to the experiment to permanent negative surface charge densities $\sigma_T = -0.07 \ldots -0.35 \, mC/m^2$ (see Methods and Supplementary Information for details). The electrode on the substrate is connected via $R_L$ to a thin Pt wire that is mounted on the top of the substrate. Upon impinging onto the solid surface ($t = 0$) at a distance $p$ from the wire the drop starts to spread and assumes a pancake structure with a pronounced rim. At a $p$-dependent time $t_o$, the drop touches the wire. After a characteristic hydrodynamic time $\tau_h$ ($\approx t_o$ for Fig. 1c), the drop reaches its maximum extent with a drop-substrate area $A(\tau_h) = A_{max}$ corresponding to a maximum spreading radius $a_{max}$ that is determined by the kinetic energy upon impact[13]. $\tau_h$ is determined by Rayleigh's inertia-capillary time scale $\sqrt{\rho a^3/\gamma}$, where $\rho$, $a$ and $\gamma$ are the density, radius, surface tension[14,15]. At long times, the drop recedes, detaches from the wire ($t = t_f$) and eventually either bounces or rolls down the slightly inclined surface (see Supplementary Information and Videos S1-S5). $t_o$ and $t_f$ can be controlled by varying the impact parameter $p$ and the slight inclination angle $\varphi = 0 \dots 30°$ of the substrate. $A(t)$ is extracted from the bottom view images (see Supplementary Section IV) and follows the expected behavior for low Ca and We numbers (Ca =5.96e-5; We ≈43, Figs. 1c, c')[13,16]. For off-wire impacts, Figs. 1 b, c, d, the simultaneously recorded current $I_R$ remains zero upon impact, increases abruptly to a peak value $I_0$ at $t = t_o$ and then relaxes to a much smaller value within a characteristic electrical time $\tau_{el}$ ($\approx 2$ ms in Fig. 1d). Eventually, the current switches sign as the drop retracts and falls abruptly to zero at $t_f$ with a finite $A(t_f)$. For on-wire impacts, Figs. 1b', c', d', the current rises continuously from zero and subsequently follows a smoother curve with much lower absolute values.

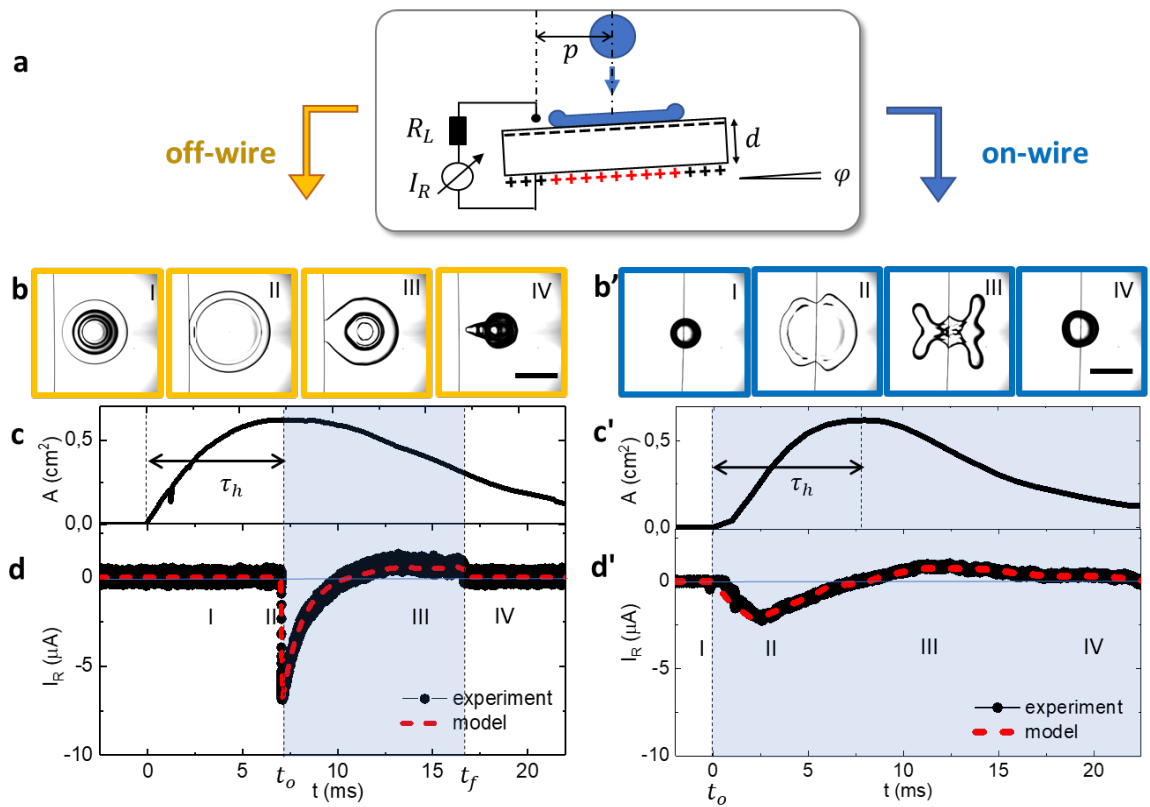

**Fig. 1| Drop impact and current measurements on negatively charged hydrophobic surface. a** experimental setup (not to scale; $d \ll$ drop radius; $\alpha = 0 \dots 30°$. **b, b'** bottom view video snapshots for various stages of impact through transparent substrate for off-wire (**b**) and on-wire (**b'**) impact.

Scale bar: 5 mm ($\varphi = 0$; $V = 30$ μL; 100 mM NaCl). **c, c'** drop-substrate contact area vs. time upon spreading an receding. **d, d'** electrical current through load resistor (**d**: $R_L = 810$ kΩ, $\sigma_T = -0.12$ mC/m²; **d'**: $R_L = 0$Ω, $\sigma_T = -0.13$ mC/m²) The current is only finite during drop-wire contact (blue region). Black symbols: experimental data; red curve: model current. (Samples are charged by Surface Charge Printing method, see Methods and Supplementary Section 1)

Exploring the dependence of the electrical current on the various parameters for off-wire impacts, we find that $I_o$ increases with decreasing $R_L$ and with increasing $\sigma_T$, while it is unaffected by the release height $h$ and thus $a_{max}$. In contrast, the electrical relaxation time increases with increasing $R_L$ and $h$, but is independent of $\sigma_T$, Fig. 2. This response represents a discharging RC circuit with an in-built initial potential difference $U_T = \sigma_T/c_d$ on the capacitor due to the trapped surface charge $\sigma_T$ and a time-dependent capacitance $C(t) = c_d A(t)$, where $c_d = \frac{\epsilon_0 \epsilon_d}{d} \approx 10^{-5} F/m^2$ ($\epsilon_0 \epsilon_d$: dielectric permittivity of the fluoropolymer layer) is the capacitance per unit area, Fig. 1a. As soon as the drop touches the wire the electrical circuit is closed and countercharge from the bottom electrode (red +'s in Fig. 1a) is transferred through $R_L$ towards the drop. (In fact, the drop gets polarized due to the trapped charge at the AFP surface forming an electric double layer (EDL) at the solid liquid interface. Likewise, an EDL forms at the wire drop interface. However, given the large specific capacitance of the EDL these two contributions, which are in series with the dielectric layer, can be neglected in the equivalent circuit.) As the current flows, the charge on the capacitor and thus the driving voltage $U(t) = U_T + q(t)/C$ decreases, where $q(t) = \int_{t_o}^{t} I_R(t')dt'$ is the total charge transferred between $t_o$ and $t$. Hence, we can write the current as

$$I_R(t)R_L = -\frac{dq}{dt}R_L = U(t) = \frac{1}{c_d}\left(\sigma_T + \frac{q(t)}{A(t)}\right) \quad (1)$$

Direct numerical integration of Equation 1 with the initial condition $q(t_o) = 0$ leads to a quantitative description of the current (red line in Figs. 1c, c') without any fit parameters. Equation 1 immediately shows that the natural unit for the current is the initial current $I_o = \sigma_T/c_d R_L$. Introducing the electrical time scale $\tau_{el} = R_L c_d A_{max}$, we can also see that $I(t)$ relaxes exponentially on this time scale for the conditions of Figs. 1b, c provided that $A \approx const.$ on the time scale $\tau_{el}$. Equation 1 also implies that for $t \gg \tau_{el}$, the charge on the capacitor follows the drop-substrate interfacial area, i.e. $q \propto A$ and hence $\dot{q} \propto \dot{A}$, explaining the current reversal for long times. Integrating the current form $t_o$ to $\tau_h$ confirms that the total transferred charge during the spreading phase is given by $Q_{tot} = \sigma_T A_{max}$ provided that $\tau_{el} \ll \tau_h$ (Supplementary Fig. S6). If the impact conditions ($\varphi, p$) are such that $A(t)$ gradually decreases to zero upon detachment from the wire, we find moreover that the charge transfer upon spreading and receding exactly compensate leaving the detaching drop with zero net charge (Supplementary Fig. S6).

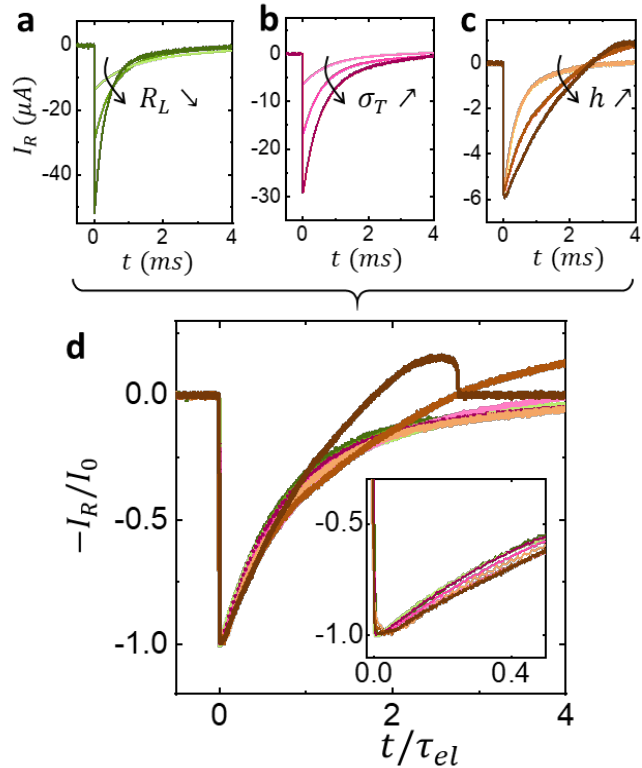

**Fig. 2 | Current response for variable impact and harvesting conditions. a,** varying $R_L$ (0.47, 0.81, 1.65 MΩ); fixed $\sigma_T = -0.35$ mC/m$^2$; h = 43mm (oxidized silicon samples); **b,** varying trapped charge density ( $\sigma_T = -0.07, -0.20, -0.35$ mC/m$^2$ ); fixed $R_L = 810$ kΩ; $h = 43$ mm (oxidized silicon samples). **c,** varying impact height ($h = 30, 90, 180$ mm); fixed $\sigma_T \approx -0.1$ mC/m$^2$; $R_L = 810$ kΩ

(transparent samples); Drop volume for **a** and **b** are 33 µL, for **c** is 17 µL. **d,** non-dimensional current response curve vs. normalized time for all data from panels **a, b, c.** Inset: zoom view for $t \ll \tau_{el}$.

So far, we focused on situations with $\tau_{el} < \tau_h$, implying that the electrical discharge is fast compared to the hydrodynamics of the impact process. The opposite situation can occur for sufficiently large $R_L$. Fig. 3 compares 'on-wire' impacts and 'off-wire' impacts for situations of $\tau_{el} \ll \tau_h$ (Fig. 3a) and $\tau_{el} > \tau_h$ (Fig. 3b). Note the occurrence of very high current peaks (up to 200µA) for off-wire impacts and $\tau_{el} \ll \tau_h$ that exceed typical currents values in the literature (nA to low µA level[17-19]) by several orders of magnitude. These high current values arise as the large charged capacitor abruptly discharges. They are enabled by our specific electrode configuration. All these different scenarios are reproduced by various limiting cases of Equation 1 (see Supplementary Information).

We can now calculate the main quantity of interest, namely the total energy dissipated in the resistor throughout the impact and rebounding process (see Methods and Supplementary Section V):

$$\Delta E = R_L \int_{t_o}^{t_f} I_R^2(t)\, dt = E_0 \ F\big(\tilde{t}_o, \tilde{t}_f, \tilde{\tau}_{el}, \{\alpha_n\}\big) \tag{2}$$

Here, $E_0 = \frac{\sigma_T^2 A_{max}}{c_d}$ is the characteristic energy of the system, and $F$ is a non-dimensional function of $\tilde{t}_o, \tilde{t}_f$, and $\tilde{\tau}_{el}$, are the corresponding times normalized by $\tau_h$ and $\{\alpha_n\}$ is a set of parameters that describes the shape of $A(t)/A_{max}$. These parameters are determined by the fluid dynamics of the impact process[16]. Note that $E_0$ is twice the electrostatic energy of the fully loaded parallel plate capacitor with $C = c_d A_{max}$ with $q_{max} = \sigma_T A_{max}$. Upon drop spreading, $q_{max}$ relaxes from its separation $d$ on the capacitor to a final separation that is given by the electric double layer thickness of a few nanometers. Upon receding, the charges are separated again back to their original configuration. Spreading thus converts electrical energy into mechanical one; receding does the opposite. However, since $\frac{E_0}{A_{max}} = \frac{\sigma_T^2}{c_d} \ll \gamma$ even for the highest $\sigma_T$ considered here, the electrical current can be harvested in both directions without taking into account the back coupling of the electrical work to the dynamics of the impact process.

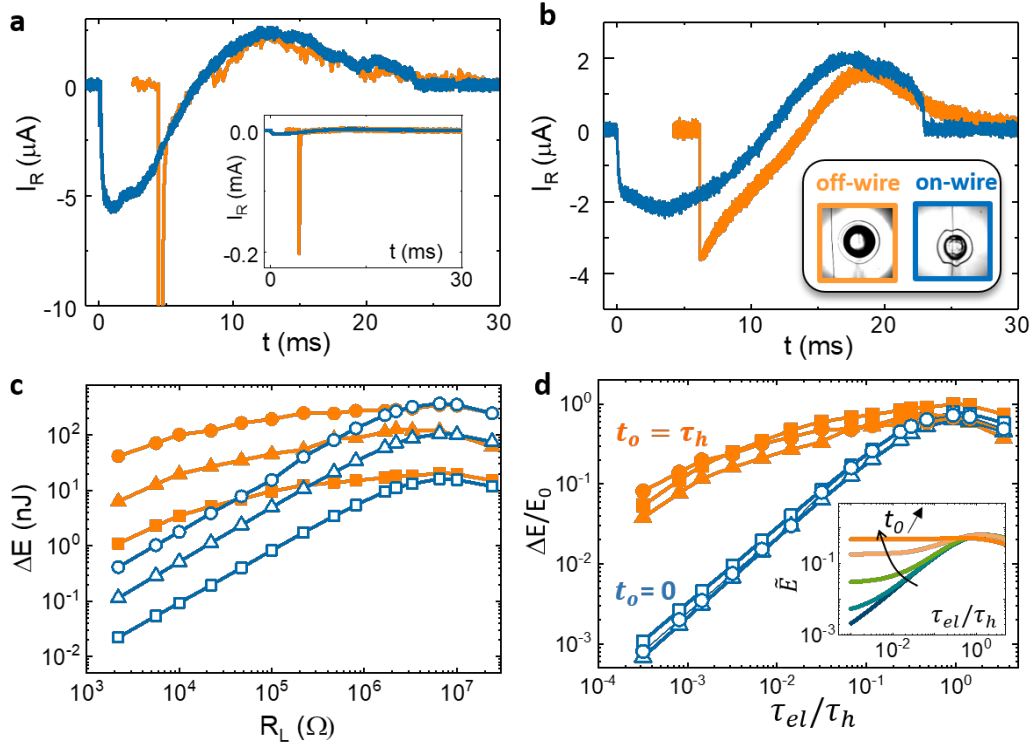

**Fig. 3 | Comparison of on-wire (blue) vs. off-wire impact (orange) for variable load resistance.** Current vs. time for **a,** $\tau_{el} \ll \tau_h$ ($R_L = 100$ kΩ, $\tau_{el} \approx 0.1\ ms, \tau_h \approx 7 ms$) and **b,** $\tau_{el} \approx \tau_h$ ($R_L = 6.5$ MΩ, $\tau_{el} = 6.3\ ms, \tau_h \approx 7 ms$). **c** Harvested energy $\Delta E$ per drop vs. load resistance for variable charge densities (squares: $\sigma_T = -0.07$ mC/m$^2$; triangles: $-0.2$ mC/m$^2$; circles: $-0.35$ mC/m$^2$. Blue: drop impact on wire; Orange: drop impact off-wire). **d** normalized energy ($\Delta E/E_0$) vs. $\tau_{el}/\tau_h$ (same data as **c**). Inset: $\Delta E/E_0$ vs. $\tau_{el}/\tau_h$ for variable $0 < t_0 < \tau_h$ based on numerical solution of Equation 1 (see Supplementary Section Ⅴ). (all data: drop volume: 33 μL; 100mM NaCl solution; $h = 43 mm$; samples are charged by EWCI method, see Methods and Supplementary Section 2)

Calculating the harvested energy from a large number of measurements for variable $\sigma_T$, $R_L$, and $t_o$, we confirm indeed the scaling as obtained in Equation 2, Fig. 3c, d. It turns out that optimum energy harvesting is only achieved for a specific value of $R_L$ corresponding to $\tau_{el} \approx \tau_h$. This is comparable to the RC circuit driven by an external alternating current, in which the energy dissipation is also maximum for driving frequencies matching the intrinsic relaxation time. While the conversion efficiency drops dramatically for $\tau_{el} \ll \tau_h$ in case of on-wire impacts, off wire impacts display a much weaker dependence. This robustness arises from the initial high current peaks under those conditions that release very quickly the entire electrostatic energy that is initially stored in the loaded capacitor. The inset of Fig. 3d shows calculated profiles of $F$ as a function of $\tau_{el}/\tau_h$ for a variety of values of $t_o/\tau_h$. For the optimum conditions of the present data set, we could harvest $\approx 0.4 \mu J$ per drop for the highest surface charge density (Fig. 3c). Given the initial gravitational energy of $14 \mu J$, this corresponds to a conversion efficiency of 2.8%, which is much higher than the previous report of 0.01% [4]. (Note that some authors use a different refence energy and thereby achieve higher apparent efficiencies [20,21])

From a materials perspective, the scaling of $E_0$ indicates that high surface charge densities, large spreading areas, and low capacitances should be sought in order to optimize the energy harvesting process. Previous attempts to achieve this goal have often suffered from poor stability and from a requirement of low conductivity[22-24]. To overcome these problems, we extended our recently developed electrowetting-based charge injection (EWCI) method [25]. By covering the electrode with a dielectrically strong $SiO_2$ layer under the AFP coating and a suitable mask to define the charged region, we can achieve stable homogeneous charge densities over surface areas of cm$^2$ by applying a voltage upon exposing the surface into water (see Methods and Supplementary Fig. S2). The present surfaces with $\sigma_T = -0.35 \text{mC/m}^2$ demonstrate robust energy harvesting for aqueous salt solutions over a wide range of concentrations, including in particular rain water and salt solutions with conductivities comparable to sea water (Fig. 4a) that did not allow for energy harvesting in conventional tribo-electric nanogenerators[5,26,27]. For the practically less relevant lowest conductivities the harvesting efficiency decreases because of the finite resistance of the liquid, which gives rise to a reduction of $I_o$ by a factor $R_L/(R_L + R_{drop})$, see inset Fig. 4a and Fig. S7, as is well-known in the electrowetting literature[1].The performance improvement of our device configuration, which we denote as charge-trapping energy nanogenerator (CT-ENG), arises from the combination of the specific electrode geometry in our experiment and the stable EWCI process that does not suffer from discharging effects upon detachment of conductive drops and exposure to multiple drops and wet environment. Long term tests of our surface demonstrate stable energy harvesting over extended periods of time without appreciable signs of degradation, Fig. 4b.

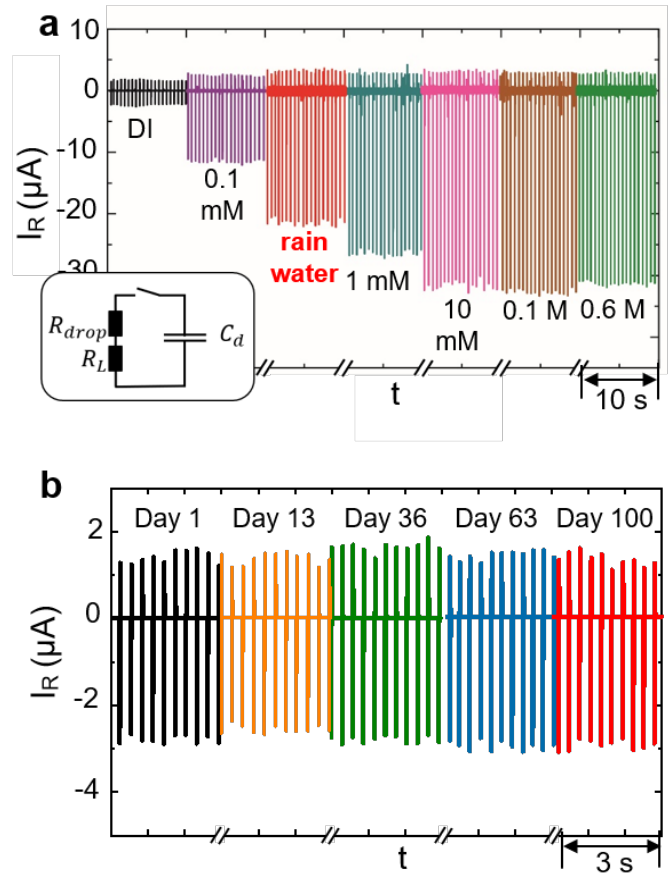

**Fig. 4 | Energy conversion and dissipation. a** Current generated from multiple water drops of variable salt concentration ($R_L = 810$ kΩ, $c_d = 1.48 \times 10^{-5} F/m^2$, $\sigma_T = -0.35 mC/m^2$). Inset: equivalent circuit indicating finite drop resistance that is at relevant low conductivity. **b** Long term stability (up to 100 days) of current response for EWCI surfaces ($R_L = 6.5$ MΩ, $c_d = 1.48 \times 10^{-5}$ F/m$^2$, $\sigma_T = -0.27$ mC/m$^2$).

Charge densities of $1 mC/m^2$ or more that can be achieved by EWCI will allow to convert more than 10% of the initial kinetic into electric energy. Even higher values are conceivable in combination with appropriate design of the capacitance $c_d$. At such levels of energy conversion the present unidirectional model is expected to reach its limits because back coupling of the electrical energy conversion will affect the drop dynamics. In terms of applications, efficiencies of 10% or higher should allow to recover appreciable amounts of energy from droplet streams with intensities comparable to bathroom showers. In that case, the typical mechanical power is of the order of $W_{mech} = 5 \times 10^5 Pa\; 10\frac{L}{min} \approx 100W$, which becomes largely available as kinetic energy of drops and usually gets lost in the sink.

In summary, our simultaneous measurements of drop spreading and electrical response provide the first quantitative description of the energy harvesting process upon drop impact and achieve a maximum energy conversion efficiency of 2.8%. We demonstrate scaling laws and identify design criteria for optimized energy conditions indicating a path towards energy conversion efficiencies beyond 10%. Our new EWCI process enables long term stable energy harvesting for a wide range of fluid compositions, including rain and sea water.

**Methods:**

**Substrate preparation and charging.** Two types of samples and charging methods were used. Transparent samples were fabricated from indium-tin-oxide (ITO)-covered glass substrates in a Class 110, ISO5 cleanroom. 800 nm thick Teflon AF1600 films (the Chemours Company, USA) were prepared by screen printer (Autech Enterprise Co., Ltd. China). Substrates were baked and annealed according to standard protocols [28] and cut into rectangular samples of 7.5 cm× 2.5 cm. Surface charges on these transparent samples are prepared by the Surface Charge Printing method, in which more than 500 drops were left to impinge on the surface prior to the experiment, leading to a constant surface charge density $\sigma_T = -0.12 \ldots -0.15$ mC/m$^2$ (see Supplementary Section Ⅰ for details).

For electrowetting-assisted charge injection (EWCI) highly doped Si wafers with a 300 nm thick oxide layer were used as substrates. The oxide layers were covered with a 1 μm thick Teflon AF 1600 layer by spin-coating (spinning speed of 1500 rpm) and annealed according to standard protocols. For the charging process, the samples were covered along the edges by a polypropylene tape acting as a mask. Subsequently, a large puddle of water was formed covered the unmasked area of the sample and a voltage of up to -400V (w.r.t. to the substrate) was applied to the drop through an immersed wire for up to 15min. This lead to surface charge densities up to $-0.35$ mC/m$^2$, as characterized by quantifying the asymmetry of the electrowetting response curve (see ref. [25,27] and Supplementary Section II ).

**Data acquisition and processing**. High-speed current measurements were performed with a fast transimpedance amplifier (HF2TA, Zurich Instrument, Switzerland) combining with a digital phosphor oscilloscope (TDS503B, Tektronix, United States). A picoammeter (Model 6487, Keithley, United States) was used for long term monitoring of sequences of drop impacts. The solid-liquid contact area during the drop impact on the surface was extracted from the images recorded with a highspeed camera (Fastcam SA5, Photron, Japan) in bottom view through the transparent ITO substrates. Standard background subtraction, filtering and binarization methods were used to quantify the in many cases non-centrosymmetric contact area $A(t)$. See Supplementary Section IV for details.

**Modeling.** We can simplify Equation 1 by introducing non-dimensional charge $x = q/q_0$, time $s = t/\tau_h$, and interfacial area $f(s) = A(\tau_h s)/A_{max}$. Using $q_0 = -\sigma_t A_{max}$, $\tau_{el} = R_L c_d A_{max}$ and $\beta = \tau_{el}/\tau_h$, we make Equation 2 dimensionless:

$$\frac{dx}{ds} + \beta^{-1}\frac{x(s)}{f(s)} = \beta^{-1}$$

For the initial condition $x(s_0) = 0$, the solution of this equation reads $x(s) = \beta^{-1}\int_{s_0}^{s} e^{-L(s,s_1)/\beta} ds_1$, where $L(s, s_1) = \int_{s_1}^{s} [f(s_2)]^{-1} ds_2$ and $s_0 < s_1 < s_2 < s$. Note that $\partial L/\partial s = 1/f(s)$. Using the experimental input for the non-dimensional area $f(s)$ (or a suitable polynomial parametrization, we integrated this equation numerically to obtain the model curves in Fig. 1. For an ideal off-wire input as in Figs. 1b, c, d the process starts at $s_0 = 1$ with $df/ds = 0$. Hence we can approximate $f(s) = 1$ and reach the exponentially decaying solutions of Fig. **2**. In the expression for the total harvested energy $\Delta E = R_L \int_{t_0}^{t_f} I^2 dt = \beta \frac{\sigma_t^2 A_{max}}{c_d}\int_{s_0}^{s_f}\left(\frac{dx}{ds}\right)^2 ds$ the integral $\beta \int_{s_0}^{s_f}\left(\frac{dx}{ds}\right)^2 ds = F(\beta, s_0, s_f, \{\alpha_n\})$ is only a function of $\beta$, $s_0$, $s_f$ and $\{\alpha_n\}$, where the parameters $\{\alpha_n\}$ describe the time evolution of $f(s)$. This leads to Equation 2 for the total energy harvested.

**Acknowledgment**

H.W. acknowledges the discussion with Lingling Shui. H.W and G.Z acknowledge support from National Key R&D Program of China (2016YFB0401501), National Natural Science Foundation of China (Grant No. 51561135014, U1501244), Program for Chang Jiang Scholars and Innovative Research Teams in Universities (No. IRT_17R40), Science and Technology Program of Guangzhou (No. 2019050001), Guangdong Provincial Key Laboratory of Optical Information Materials and Technology (No. 2017B030301007), MOE International Laboratory for Optical Information Technologies and the 111 Project.

**Author contributions**

H.W., G. Z. and F. M. conceived the project. H. W. and F. M. designed the experiment. H. W. and N. M. performed the experiment. H. W., N. M.and F. M. analyzed the experimental data. H. W., D. van den

E. and F. M. built the model. H. W. and F. M. wrote the draft. All authors contributed to revising the manuscript.

**Competing interests**

The authors declare no competing interests.

**Supplementary information**

Supplementary Section I- VI and Video S1-S6.

# Supplementary information

# Electric generation from drops impacting onto charged surfaces

Hao Wu[1,2,3*], Niels Mendel[3], Dirk van den Ende[3], Guofu Zhou[1,2,4*], Frieder Mugele[3*]

[1] Guangdong Provincial Key Laboratory of Optical Information Materials and Technology & Institute of Electronic Paper Displays, South China Academy of Advanced Optoelectronics, South China Normal University, Guangzhou 510006, P. R. China

[2] National Center for International Research on Green Optoelectronics, South China Normal University, Guangzhou 510006, P. R. China

[3] Physics of Complex Fluids, Faculty of Science and Technology, MESA+ Institute for Nanotechnology, University of Twente, Enschede 7500AE, the Netherlands

[4] Shenzhen Guohua Optoelectronics Tech. Co. Ltd., Shenzhen 518110, P. R. China

*Emails: haowu.ut@gmail.com (H.W.); guofu.zhou@m.scnu.edu.cn (G.Z.); f.mugele@utwente.nl (F.M.)

## I. Charging transparent substrates by surface charge printing

In order to record the evolution of the drop-substrate contact area, we need high-speed imaging through the bottom of the substrates. Therefore, transparent substrates are required. As reported previously, surface charges can be generated from a water drop impacting on/contacting with hydrophobic surfaces[1-3]. For instance, Q. Sun et al. used surface charge printing (SCP) for programmed droplet transport[1]. In this work, we use SCP to generate a surface charge distribution on a fluoropolymer surface. The schematic is shown in Fig. S1a. A series of droplets with volume of 33 μL are released from a height of around 5 cm and impacts on a fixed spot on the hydrophobic surface. This results in a surface charge at the impact spot. According to our observation, the surface charge increases with every drop impact. The surface charge density reaches a plateau after around 500 drops (Fig. S1b). This surface charge saturation phenomenon has also been observed for the case of drop sliding on a hydrophobic perfluoro octadecyltrichlorosilane (PFOTS) surface[3]. With the SCP method, surface charges of approximately 0.15 mC/m$^2$ can be generated (depending on drop height and substrate conditions).

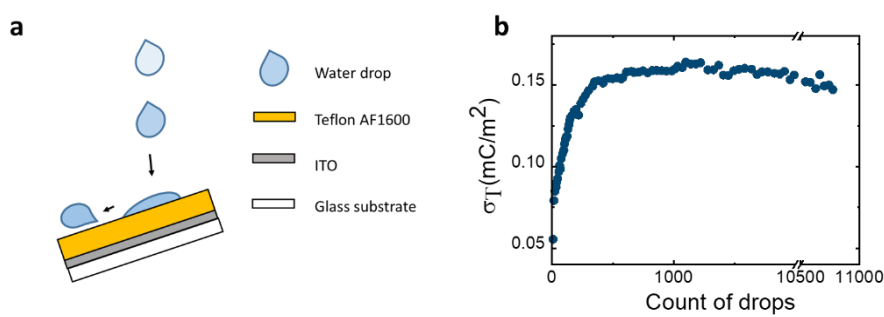

**Fig. S1 a**, schematic of surface charge produced by surface charge printing(SCP) method. **b**, surface charge density depending on the count number of impacting water drops.

## II. Charging Si/SiO$_2$/Teflon substrates by improved electrowetting-assisted charge injection (EWCI)

In order to verify the proposed physical model and to enhance the performance of the Charge Trapping Electrical Nanogenerator (CT-ENG), substrates with a stable and high surface charge density are required. For this purpose, ElectroWetting-assisted Charge Injection (EWCI) method is applied.

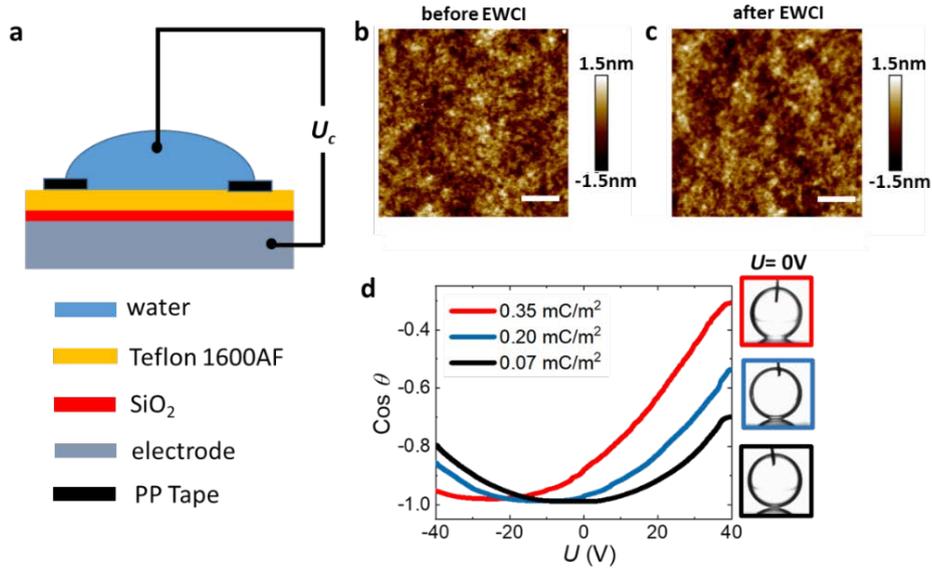

**Fig. S2 a**, schematic of the enhanced Electrowetting-assisted Charge Injection (EWCI) process (thickness of SiO$_2$: 300nm; thickness of Teflon AF1600: 1μm). Atomic force microscopy (AFM) images of surfaces **b**, before and **c**, after EWCI process. Scale bar: 200 nm. **d**, surface charge densities on three charged surfaces measured by Electrowetting probe. Charge injection conditions for these three samples are, red: -400V for 15 min; blue: -300V for 15 min; black: non-charged. Insets are the drop images under applied voltage $U = 0$V. The frame color indicates the surface charge density: red for $-0.35$ mC/m$^2$, blue for $-0.20$ mC/m$^2$ and black for $-0.07$ mC/m$^2$.

According to Chapter 4, charges can be injected at the three phase contact line (TPCL) region on fluoropolymer-water interfaces during electrowetting due to the locally enhanced electric field. To deposit a homogeneous charge distribution over the full contact area, a relatively high and more homogeneous electric field should be applied. To this purpose, we protect the substrate near the TPCL region by polypropylene (PP) tape and introduce a 300 nm thin thermally grown SiO$_2$ layer as a dielectric layer underneath the fluoropolymer (Teflon AF1600) layer. The dielectric strength of the thermally grown SiO$_2$ is higher than 1000 V/μm, which is much higher than for Teflon AF (20 - 150 V/μm)[4,5]. By simply placing a 300 nm SiO$_2$ layer underneath the 1 μm fluoropolymer film, a potential of 400 V can be applied to the combined film (using deionized water) without damage. Consequently, a high electric field can be applied over a large arear of the dielectric layer. Fig. S2 shows the schematic of this improved EWCI process. It also shows that the surface topography does not change under EWCI, consistent with the results of Chapter 4. After charging the surface in this way, the water is removed from the fluoropolymer surface and the surface charge densities can be tested by electrowetting (EW), as shown in Fig. S2d. For a neutral surface, the EW response curve, i.e. $\cos\theta$, where $\theta$ is the contact angle, versus applied voltage $U$, is symmetric around $U = 0$. When the hydrophobic surface has been charged, the symmetry axis will be shifted to $U = U_T$. From this shift $U_T$, the surface charge density can be calculated as $\sigma_T = c_d U_T$, where the $c_d$ is the capacitance of the dielectric layer per area [6]. For a pristine surface, a spontaneous surface charge density of -0.07 mC/m² has been found. By applying -300V or – 400V for 15 minutes a charge density of respectively -0.20 mC/m² and -0.35 mC/m² is achieved.

### III. Charge transfer process

As shown in Fig. 1, the process of a drop impacting on a solid surface can be divided into 4 stages: I) the drop impacts and spreads on the charged surface; II) the drop reaches its maximum spreading and touches the conductive (Pt) wire; III) the drop contracts; IV) the drop detaches from the wire. Here, we discuss the charge transfer during these stages, using the schematics and the equivalent circuits that are shown in Fig. S3.

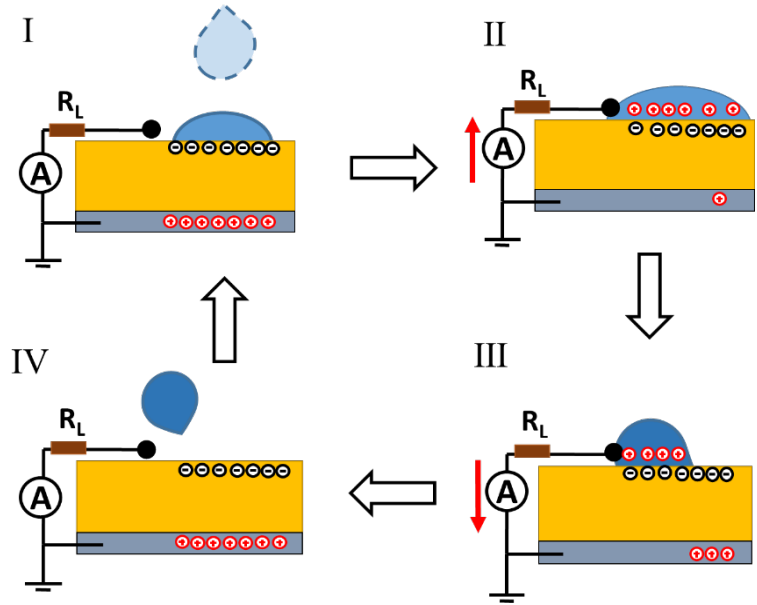

**Fig. S3** Schematic of the four stages of a drop impacting on a charged surface: I) drop impacting and spreading on the charged surface; II) drop reaching its maximum spreading and touching the conductive (Pt) wire; III) drop contracting; IV) drop detaching the wire.

I) Before the drop touches the wire, all counter charges are at the bottom electrode. Since the switch in the equivalent circuit is open, there flows no current through the load resistor $R_L$.

II) Because the capacitance of electric double layer($C_{EDL}$) at the liquid-solid interface is much larger than the dielectric capacitance ($C_{Diel}$), counter charge tends to migrate from the bottom electrode through $R_L$ to the liquid-solid interface, when the drop touches the wire. The amount of charge transferred between the two electrodes depends on both the hydrodynamic $\tau_h$ and electric timescale $\tau_{el}$ we discussed in the main text. When the resistance is small, $\tau_{el} \ll \tau_h$, and the local counter charge will be transferred to the EDL.

III) After reaching its maximum spreading, the drop starts to contract, and the liquid-solid contact area $A(t)$ decreases. As shown in Fig. S3-III, the counter charge flows back to the bottom electrode during this stage, leading to a positive current through $R_L$. In the case $\tau_{el} \ll \tau_h$, the current $I_R$ is dominated by the drop dynamics, while when $\tau_{el}$ is comparable or larger than $\tau_h$, $I_R$ is determined by both the drop dynamics and the RC circuit response.

IV) When the drop detaches from the wire and bounces off or slides downhill (depending on whether the surface is tilted), the current through the load resistor becomes zero.

### IV. Extracting the liquid-substrate contact area A(t)

The impacting drop is observed through the substrate using a microscope in reflection mode. Although it is more vividly to observe the drop impact from aside, one can only observe the outer profile of the drop, while the liquid-solid contact area is hard to quantify. To determine the liquid-solid contact area $A(t)$ correctly, we process the images observed in a reflection mode.

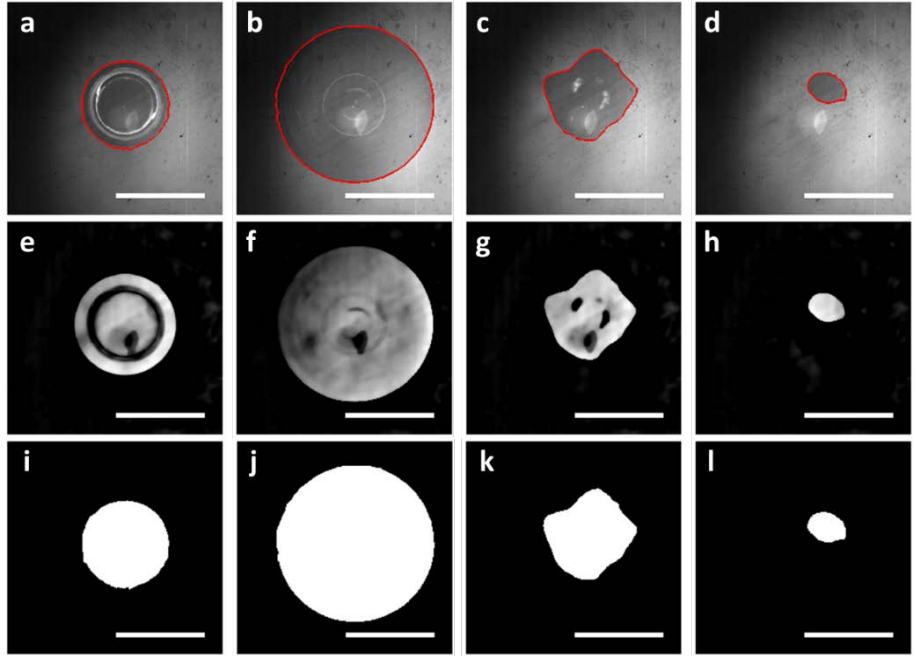

**Fig. S4 a-d,** non-processed images (extracted liquid-solid interfacial area is marked in red, see Video S6); **e-h**, images after background subtraction and median filtering (steps 1 and 2); **i-l**, binary masks of the extracted liquid-solid interfacial area. Scale bar: 5mm.

Images are recorded using a high speed camera (Fastcam SA5, Photron, Tokyo, Japan) at 10,000 fps and a microscope (Eclipse Ti, Nikon, Tokyo, Japan). Due to limited contrast with the background, the liquid-solid contact area could not be extracted by simple thresholding techniques (Fig. S4a-d). Therefore, we process the images in several steps (using the scikit-image library[7], Fig. S4e-l):

1) The background is removed by subtracting a Gaussian blur of the image (kernel: 40px disk), and subtracting the first recorded image in the sequence (with zero liquid-solid interfacial area). Subtracting a Gaussian blur with a large kernel removes the background gradient. Subtracting the initial image reduces the noise in the images considerably (e.g. noise from pollution of the backside of the substrate).
2) A median filter (kernel: 10px disk) is applied to the images. Other than by applying a standard Gaussian blur to smoothen the image, a median filter smoothens the image while preserving sharp edges. (Figs. S4.4 e-h).
3) Edges are detected by applying a 3x3 Sobel filter. The output image is binarized by applying a (global) threshold.
4. Morphological transformations are applied to remove small objects and fill holes from the binary mask(Figs. S4i-l). The number of pixels on the resulting mask is counted and converted using a predetermined scaling factor.

## V. Modeling the electric behavior

The governing equation for the charge flowing through the load resistor is given by Eq. 1 :

$$-\frac{dq}{dt} = I(t) = \frac{U(t)}{R_L} = \frac{\sigma_t A(t) + q(t)}{R_L c_d A(t)} \tag{S1}$$

Where $t_0 < t < t_f$ and $q(t_0) = 0$. This equation can be rewritten as:

$$\frac{dq}{dt} + \frac{q(t)}{R_L c_d A(t)} = \frac{-\sigma_t}{R_L c_d} \tag{S2}$$

Defining $x = q/q_0$, $s = t/\tau_h$, $f(s) = A(\tau_h s)/A_{max}$, $q_0 = -\sigma_t A_{max}$, $\tau_{el} = R_L c_d A_{max}$ and $\beta = \tau_{el}/\tau_h$, we make Eq. S2 dimensionless:

$$\frac{dx}{ds} + \beta^{-1}\frac{x(s)}{f(s)} = \beta^{-1} \tag{S3}$$

By inspection we observe that the solution of Eq. S3, with boundary condition $x(s_0) = 0$, is given by:

$$x(s, \beta, s_o) = \beta^{-1} \int_{s_0}^{s} e^{-\beta^{-1} L(s, s_1)} ds_1 \tag{S4}$$

where $L(s, s_1) = \int_{s_1}^{s}[f(s_2)]^{-1} ds_2$ and $s_0 < s_1 < s_2 < s$. Note that $\partial L/\partial s = 1/f(s)$. Eq. S4 can be solved numerically, once one has an expression or parametrization for $f(s)$.

From Eq. S3 we observe that for $s = s_0$ the derivative is given by $dx/ds = \beta^{-1}$ unless $f(s_0) = 0$. In that case we obtain for the derivative, using l'Hopital's rule: $dx/ds = \beta^{-1} f'(s_0)/(f'(s_0) + \beta^{-1})$, where $f'(s)$ is the derivative of $f(s)$.

In dimensional form we obtain for the charge $q$ and current $I = -dq/dt$ as a function of time $t$:

$$q(t) = -\sigma_t A_{max} x(s), \qquad I(t) = \frac{\sigma_t A_{max}}{\tau_h} \left(\frac{dx}{ds}\right) \tag{S5}$$

for $t/\tau_h = s > s_o = t_o/\tau_h$. Thus for $t = t_0$ we get in case $A(t_0) \neq 0$ (off-wire):

$$I_0 = \frac{\sigma_t A_{max}}{\tau_{el}} = \frac{\sigma_t}{R_L c_d} \tag{S6}$$

and in case $A(t_0) = 0$ (on-wire):

$$I_0 = \frac{\sigma_t A'(t_0)}{1 + R_L c_d A'(t_0)} \simeq \frac{\sigma_t}{R_L c_d + \tau_h/A_{max}} \tag{S7}$$

Here we use the notation $A'(t) = dA/dt$. At impact we estimate $A'(0) \approx A_{max}/\tau_h$. Therefor, the off-wire current can be much larger than the on-wire current. The excess current decays exponentially with a short decay time because at $t = t_0$:

$$\left(\frac{d^2x}{ds^2}\right)_{s_0} = \frac{-\beta^{-1}}{f(s_0)} \left(\frac{dx}{ds}\right)_{s_0} \tag{S8}$$

So the initial slope of $I(t)$ is given by $(dI/dt)_{t_0} = -I_0/\tau_{el}$.

If the process starts at $s_0 = 1$, where $df/ds = 0$, we can approximate Eq. 3 for $1 < s \ll 2$ as:

$$\frac{dx}{ds} + \beta^{-1} x(s) = \beta^{-1} \tag{S3a}$$

Which has as solution $x(s) = 1 - e^{-\beta^{-1}(s-s_0)}$ and $dx/ds = \beta^{-1} e^{-\beta^{-1}(s-s_0)}$. In dimensional form last equation reads:

$$I(t) = \frac{\sigma_t A_{max}}{\tau_{el}} e^{-(t-t_0)/\tau_{el}} \tag{S3b}$$

The total harvested energy is given by:

$$\Delta E = R_L \int_{t_0}^{\infty} I^2 dt = \beta \frac{\sigma_t^2 A_{max}}{c_d} \int_{s_0}^{s_f} \left(\frac{dx}{ds}\right)^2 ds \tag{S9}$$

Because the integral

$$\beta \int_{s_0}^{s_f} \left(\frac{dx}{ds}\right)^2 ds = F(\beta, s_0, s_f, \{\alpha_n\}) \tag{S10}$$

is just a function of $\beta = \tau_h/\tau_{el}$, $s_0$, $s_f$ and $\{\alpha_n\}$, where the parameters $\{\alpha_n\}$ describe the time evolution of $A(t)/A_{max}$, we can write the harvested energy as:

$$\Delta E = \frac{\sigma_t^2 A_{max}}{c_d} g(\beta, s_0, s_f, \{\alpha_n\}) \qquad (S11)$$

We can tune $\beta$ and $s_0$ such that $\Delta E$ is optimal, i.e. $\partial F/\partial \beta = \partial F/\partial s_0 = 0$.

With this model in mind we now explain several observations done during this study.

**Initial current**

According to the model, see Eq. S6, the initial current value when the drop touches the wire at a "off-wire" mode can be calculated as $I_0 = \sigma_T/c_d R_L$, depending not only on the load resistance, but also on the surface charge density $\sigma_T$. This prediction is also confirmed by the experimental results shown in Fig. S5. The currents measured in Figs. S5 a and b, using substrates with a charge density of $-0.07$, $-0.20$ and $-0.35$ mC/m², directly show that a higher $\sigma_T$ results in a higher $I_0$. With a load resistance of 47 kΩ, milliamp level initial currents can be achieved. But the low resistance also leads to a shorter electrical relaxation time ($\tau_{el} = R_L c_d A_{max}$). The peak width of the current for $R_L = 47$ kΩ is below millisecond, which is much narrower than for $R_L = 6.5$ MΩ (inset of Fig. S5b). $I_0$ and the corresponding initial power ($P_0$) are shown in Fig. S5d for a wide range of $R_L$ values. Taking into account a 10 ~20 kΩ internal resistance introduced by the detection electronics ($R_{in}$), we fit the $I_0(R_L)$ curve with $I_0 = \sigma_T/c_d(R_L + R_{in})$ varying $\sigma_T$. The fitted value of $\sigma_T$ is consistent with the EW measurements, as discussed in Supl. Sect. II.

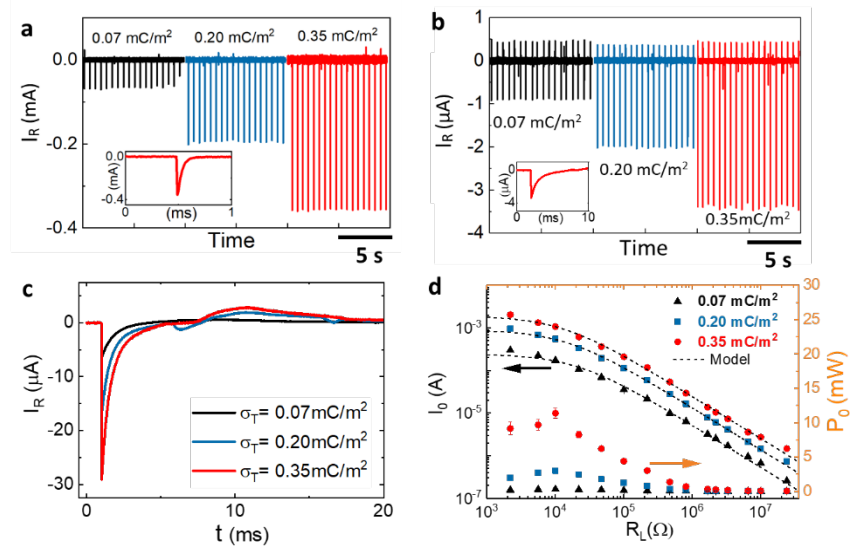

**Fig. S5** Current $I_R(t)$ for charge densities $\sigma_T$ of $-0.07$, $-0.20$ and $-0.35$ mC/m² with load resistor **a,** $R_L = 47$ kΩ and **b,** $R_L = 6.5$ MΩ. Insets show $I_R(t)$ for $\sigma_T = -0.35$ mC/m² . **c,** $I_R$ from a single drop impact for different $\sigma_T$ values ($R_L = 810$ kΩ). **d,** Initial current $I_0$ and power $P_0$ versus load resistance $R_L$. The dashed line shows the best fitting curve $I_0(R_L) = \sigma_T/R_L c_d$. The substrates are the same as in Fig. S2.

**Conservation of transferred charge**

In the situation that the drop impacts on a flat surface in a "off-wire" mode, there is always a finite contacting area when the drop detaches from the wire, which cannot be neglected (see Fig. S1). However, after reaching the maximum spreading, the impacting drop will contract and bounce off (Fig. S6a). As a result, the finite area when the drop detaches from the surface could be very small (the red circle in the Fig. S6a). Correspondingly, almost all the charge transferred to the EDL is transferred back to the bottom electrode as the drop contracts and bounces off the surface. As shown in Fig. S6b, the integral over the negative part of the $I_R$ and the positive part are identical.

When $R_L$ is small, and $\tau_{el}$ is short, all charge in the EDL near the liquid-substrate contact area can transfer back and forth between the bottom electrode and the EDL during drop spreading and contraction. However, when $R_L$ is large and $\tau_{el}$ comparable with $\tau_h$, charge transfer is partially blocked by the large $R_L$ in the circuit. In Fig. S6c, is also shown that the amount of transferred charge $q$ is higher for samples with a larger surface charge density. The maximum $q$ for the samples with a surface charge of $-0.07$, $-0.20$ and $-0.35$ mC/$m^2$ are ~23 nC, 12 nC and 5 nC, respectivily. Considering the maximum spreading area of a 33 µL drop released from a 4.3 cm height is around 0.62 cm², the transferred charge density of these three samples are approximately 0.08 mC/$m^2$, 0.24 mC/$m^2$ and 0.37 mC/$m^2$. These results are consistent with the EW results in Suppl. Sect. II.

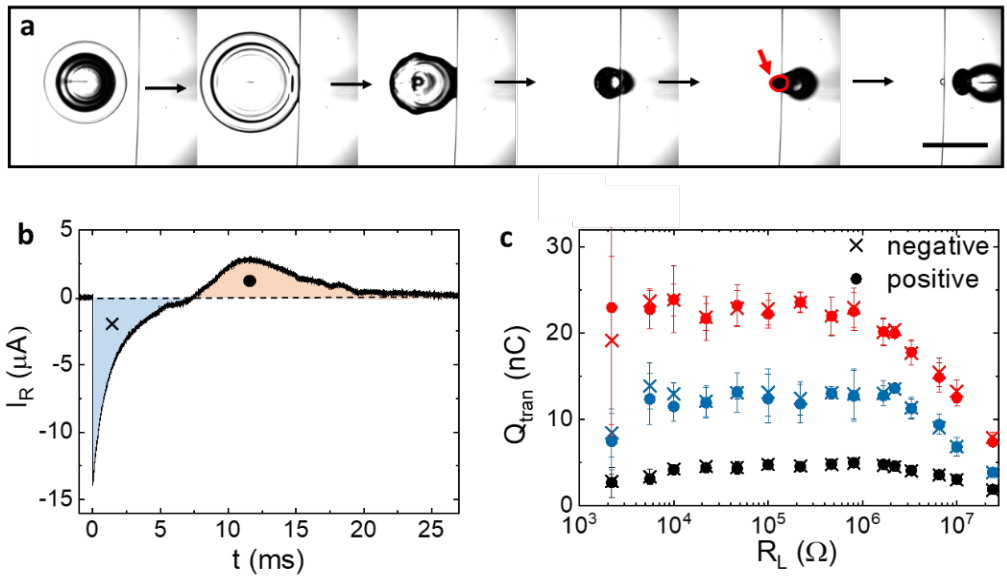

**Fig. S6 a**, bottom view through a tilted transparent substrate of a drop (volume: 33µL; liquid: 100 mM NaCl solution) impacting on a hydrophobic surface (Teflon AF1600). **b**, generated current ($I_R$) from a 33 uL drop impacting on a surface with charge density ($\sigma_T$) of -0.35 mC/$m^2$. **c**, the amount of transferred charges q with samples of $\sigma_T$ =-0.07 mC/$m^2$ (black) , -0.20 mC/$m^2$ (blue) and -0.35 mC/$m^2$ (red) depending on $R_L$. The "x" represents current direction from the bottom to the EDL at liquid-solid interface (blue part in **b**), " • " represents the reverse current direction (orange part in **b**) .

### Dependence of conductivity of the drop

In a typical charge trapping electrical nanogenerator (CT-ENG), the resistance of the drop can be neglected when the conductivity of the liquid is relatively high, because the resistance of the water drop $R_{drop} \ll R_L$. The current is calculated as $I_R(t) = U(t)/R_L$, where the $U(t)$ is the potential difference across the dielectric capacitor.

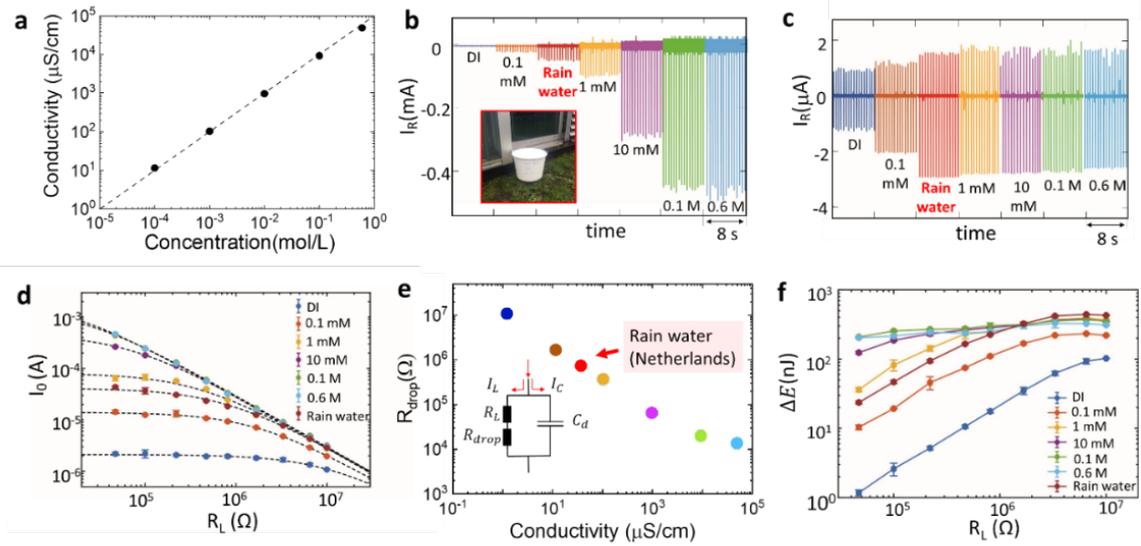

**Fig. S7 a**, conductivity depending on the salt (NaCl) concentration. Current ($I_R$) generated from multiple falling drops with various salt concentration, as well as the rain water from the campus of University of Twente (Enschede, the Netherlands) with load resistance of **b**, 47 kΩ and **c**, 10 MΩ. Inset of **b** shows rain water collection. **d**, the characteristic current value $I_0$ depending on load resistance with various salt concentration of drops. **e**, calculated resistance induced by drops ($R_{drop}$) depending on drop conductivity. **f**, generated energy ($\Delta E$) depending on load resistance with various salt concentration of drops.

However, when a low conductive liquid is used for the falling drop, $R_{drop}$ can be comparable or even larger than $R_L$, therefore $R_{drop}$ cannot be ignored and the current is $I_R(t) = U(t)/(R_L + R_{drop})$. The initial current at the moment the drop touches the Pt wire is $I_0 = \sigma_T/c_d(R_L + R_{drop})$.
NaCl solutions with various concentrations were prepared to investigate the effect of the conductivity of the liquid. The linear relationship of the conductivity and the concentration of NaCl solutions is shown in Fig. S7a. The current curves for various conductivities of the drop are shown in Figs. S7b and c. For a low load resistance of 47 kΩ, $I_0$ increases with increasing conductivity over a large range. $I_0$ becomes constant when the NaCl concentration reaches 0.1 M and $R_{drop}$ becomes negligible compared to $R_L$. For a relatively high load resistance of 10 MΩ, $R_{drop}$ hardly affects $I_0$, and only when the NaCl concentration is below 1 mM, the $I_0$ value can be reduced. In Fig. S7d $I_0$ versus $R_L$ is shown for various salt concentrations. By varying $R_{drop}$ we obtain the best fitting curve $I_0(R_L) = \sigma_t / c_d(R_L + R_{drop})$. The values found for $R_{drop}$ are inversely proportional to the liquid conductivity. Low conductive drops introduce a certain resistance to the circuit, and as such $R_{drop}$ consumes energy. Consequently, the energy $\Delta E$ generated in the load resistor will be reduced by the low drop conductivity (Fig. S7f).

**Calculating the harvested energy ΔE**
Using Eqs. S4 and S11, and a typical area profile $A = A_{max}f(s)$ of a drop that impacts on a hydrophobic surface (Fig. S8a), we calculate the energy $\Delta E = E_0\, g(\beta, s_0, s_f, \{\alpha_n\})$, where $g$ depends on the load resistance, via $\tau_{el} = \tau_h/\beta = R_L c_d A_{max}$. The results have been shown in Fig. S8 b. From this calculation we learn that for $\tau_{el} < \tau_h$ most energy is harvested when the drop touches the Pt wire at maximal spreading.

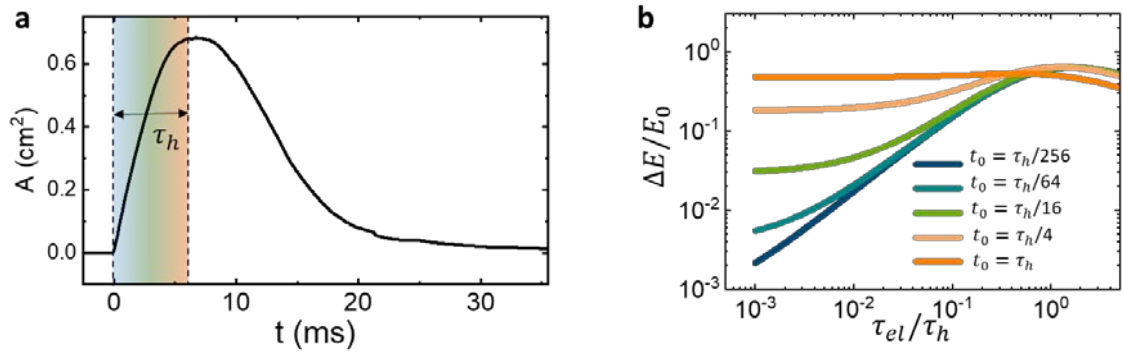

**Fig. S8 a,** Typical time profile of the contacting area $A(t)$ for a water drop impacting on a hydrophobic surface. **b,** Calculated $\Delta E/E_0$ versus $\tau_{el}/\tau_h$ for various drop impact times ($0 < t_0 \leq \tau_h$).

## VI. Description of the Supplementary Videos

**Video 1:** Drops impact on a tilted charged AFP surface. (Also the setup of the CT-ENG)

**Video 2:** Bottom view of drops off-wire impact on a flat AFP surface.

**Video 3:** Bottom view of drops on-wire impact on a flat AFP surface.

**Video 4:** Bottom view of drops off-wire impact on a tilted AFP surface ($\varphi=15°$).

**Video 5:** Bottom view of drops on-wire impact on a flat AFP surface. ($\varphi=15°$)

**Video 6:** Extracting the water-solid contacting area. Scale bar: 5mm